\documentclass[12pt,fleqn]{article}
\usepackage[active]{srcltx}
\usepackage{graphicx}
\usepackage{amsmath}
\usepackage{bbold}
\usepackage{eufrak}
\usepackage{caption}
\usepackage{slashed}
\title{\large A Master Equation Approach to the `$3+1$' Dirac Equation}
\author{\large Keith A. Earle, \\ 
Physics Department, University at Albany\\
1400 Washington Ave, Albany NY 12222}

\begin{document}

%
%
\newcommand\Dlmk[4]
       {\mbox {$\mathcal{D}^{#1}_{#2\,#3}(#4)$}}
\newcommand\dlmk[4]
       {\mbox {$d^{#1}_{#2\,#3}(#4)$}}
\newcommand\WthreeJ[6]
       {\mbox {$\left(\begin{array}{ccc} #1 & #2 & #3 \\ #4 & #5 & #6
\end{array}\right)$}}
\newcommand\WsixJ[6]
       {\mbox {$\left\{\begin{array}{ccc} #1 & #2 & #3 \\ #4 & #5 & #6
\end{array}\right\}$}}
\newcommand\WnineJ[9]
       {\mbox {$\left\{\begin{array}{ccc} #1 & #2 & #3 \\ #4 & #5 & #6 \\ #7 & #8 & #9
\end{array}\right\}$}}
\newcommand\VCfourJ[9]
       {\mbox {$\left[\begin{array}{ccc} #1 & #2 & #3 \\ #4 & #5 & #6 \\ #7 & #8 & #9
\end{array}\right]$}}
\newcommand\CG[6]
       {\mbox {$\left( #1 \, #2 \, {#4} \, {#5} \right| \left. #3 \, {#6}\right)$}}
\newcommand\QO[2]
			 {\mbox {$Q^{[#1]}_{#2}$}}

\newcommand\trace{\mbox {\textrm{Tr}}}
\newcommand\superoperator[1]{\mbox {$\mathbf{\mathcal{#1}}$}}
\newcommand\flip{\superoperator{F}}
\newcommand\CC{\superoperator{C}}
\newcommand\Yop{\superoperator{Y}}
\newcommand\Liouv{\superoperator{L}}
\newcommand\Ident{\superoperator{I}}
\newcommand\FP{\mbox {$\mathbf{\Gamma}$}}
\newcommand\SFP{\mbox {$\tilde{\FP}$}}

%
%
%
\newcommand\bra[1]
           {\mbox {$\langle\>#1\>\vert$}}
\newcommand\ket[1]
           {\mbox {$\vert\>#1\>\rangle$}}
\newcommand\braket[2]
           {\mbox {$\langle\>#1\>\vert\>#2\>\rangle$}}
\newcommand\matrixelement[3]
           {\mbox {$\langle\>#1\>\vert\>#2\>\vert\>#3\>\rangle$}}
\newcommand\reducedME[3]
					 {\mbox {$\langle\>#1\>\Vert\>#2\>\Vert\>#3\>\rangle$}}
\newcommand\projector[2]
           {\mbox {$\vert\>#1\>\rangle\langle\>#2\>\vert$}}
\newcommand\CME[6]
           {\mbox {$C_{#1,#2\, #3,#4}^{(#5)#6}$}}
\newcommand\XME[4]
           {\mbox {$X_{#1,#2}^{(#3)#4}$}}
%
%
%
%
\newcommand\llangle{\mbox{{$\langle\langle$}}}
\newcommand\rrangle{\mbox{{$\rangle\rangle$}}}
\newcommand\sbra[1]
           {\mbox {$\llangle\>#1\>\vert$}}
\newcommand\sket[1]
           {\mbox {$\vert\>#1\>\rrangle$}}
\newcommand\sbraket[2]
           {\mbox {$\llangle\>#1\>\vert\>#2\>\rrangle$}}
\newcommand\smatrixelement[3]
           {\mbox {$\llangle\>#1\>\vert\>#2\>\vert\>#3\>\rrangle$}}
\newcommand\sprojector[2]
           {\mbox {$\vert\>#1\>\rrangle\llangle\>#2\>\vert$}}
%
%
%
%
\newcommand\mbra[1]
           {\mbox {$((\>#1\>\vert$}}
\newcommand\mket[1]
           {\mbox {$\vert\>#1\>))$}}
\newcommand\mbraket[2]
           {\mbox {$((\>#1\>\vert\>#2\>))$}}
\newcommand\mmatrixelement[3]
           {\mbox {$((\>#1\>\vert\>#2\>\vert\>#3\>))$}}
\newcommand\mprojector[2]
           {\mbox {$\vert\>#1\>))((\>#2\>\vert$}}

\maketitle

\begin{abstract}
A derivation of the Dirac equation in `$3+1$' dimensions is presented based on a master equation approach
originally developed for the `$1+1$' problem by McKeon and Ord. The method of derivation presented here
suggests a mechanism by which the work of Knuth and Bahrenyi on causal sets may be extended to a 
derivation of the Dirac equation in the context of an inference problem.
\end{abstract}

\section{Introduction}

The Feynman Checkerboard (or Chessboard) problem\cite{FH:65,S:05} is a model from which the Dirac
Equation\cite{BD:64} in `$1+1$' dimensions may be derived. Feynman's version of the problem was first
published in his textbook on path integral methods\cite{FH:65}. A combinatoric solution was published
some years later\cite{JS:84}. That work, along with other combinatoric approaches, are critically
reviewed and corrected elsewhere\cite{E:10}. In addition to combinatoric methods of solution,
Gersch\cite{G:81} published a solution based on the correspondence between the Feynman Checkerboard model
and the one-dimensional Ising model. This approach was developed by Ord and coworkers in a series of
papers\cite{O:92a,O:92b,MO:92,OM:93}. In particular, the paper by McKeon and Ord\cite{MO:92} introduced
a master equation approach for the solution of the `$1+1$' version of the problem which considered
contributions from propagation events forwards and backwards in time. By imposing a causality constraint,
the `$1+1$' Dirac equation emerged in a fairly straightforward fashion. The impressive variety of 
methods available for solving the Checkerboard problem is complemented by the range of systems to which
the basic model may be applied. As an example, work by Kholodenko\cite{K:90,K:91} showed how the
available approaches could be extended to, \textit{e.g.}, polymer dynamics and heterotic strings, arenas
seemingly far removed from the original provenance of a textbook exercise.

One shortcoming of the `$1+1$' Feynman Checkerboard model is that it does not account for spin. The
history of spin within the context of a path integral approach is a checkered one, to coin a phrase.
One possible approach was spearheaded by Schulman\cite{S:68}. In some sense, the success of the `$1+1$ 
model is due to the observation that spin seems not to exist as an independent concept in one spatial 
dimension\cite{S:05}. Extensions to three dimensions have hitherto all seemed to founder on the 
difficulty of accommodating the required $\mbox{\boldmath $\sigma$}\cdot\mathbf{p}$ operator. The work
presented here indicates one possible method for accomplishing this.

An intriguing application of this approach may be found in an extension of the investigations of Knuth
and Bahrenyi\cite{KB:09} on causal sets, or posets, to a derivation of special relativity. The duality
property of posets\cite{DP:02} suggested to the author that the forward and backward master equation
approach exploited by McKeon and Ord\cite{MO:92} might offer some insights into how to extend the
poset approach of Knuth and Bahrenyi\cite{KB:09} to a derivation of the `$3+1$' Dirac Equation. This is
work in progress. The advantage of the master equation approach is that explicit expressions for
real-valued quantities may be obtained at each time step, facilitating comparison to the poset approach.
Complex amplitudes appear naturally as a result of a discrete (invertible) Fourier transform from time
and space variables to a momentum space representation. At each stage, invertible, unitary
transformations allow one to `follow the breadcrumbs' of a `$3+1$' Dirac equation back to the original,
real-valued master equation. These assertions will be validated in the sequel.

\section{Master Equation Approach}

Define a transition rate $a$ for changing direction on a line. It will be seen that this transition
rate may be interpreted as a particle mass. The transition probability in time
$\Delta t$ will then depend on $a\Delta t$. McKeon and Ord\cite{MO:92} then write down a master
equation for the probability amplitude of heading towards increasing values $+$ or decreasing values
$-$ of $x$
\begin{displaymath}
 P_{\pm}(x,t+\Delta t) = (1-a\Delta t)P_{\pm}(x\mp\Delta x,t)+a\Delta tP_{\mp}(x\pm\Delta x,t).
\end{displaymath}
As McKeon and Ord showed\cite{MO:92}, the master equation for $P_{\pm}(x,t)$ may be iterated to obtain
\begin{displaymath}
 P_{+}(x,t) = \sum_{\mbox{paths}}(1-a\Delta t)^{n-R}(a\Delta t)^R
\end{displaymath}
where $n$ is the number of steps, and $R$ is the number of reversals. Setting $a\Delta t=i\epsilon$ and
$(1-a\Delta t)\approx 1$ reproduces Feynman's expression\cite{FH:65} for the probability amplitude.
It is not clear from the analysis, however, that the substitution $a\Delta t=i\epsilon$ is a reasonable
thing to do, although it does give the right answer.

In order to address this issue, McKeon and Ord\cite{MO:92} developed a more versatile approach based on
coupled master equations, allowing for the possibility of forward and backward propagation subject to a 
causality constraint. In a notation adapted from Ord and McKeon\cite{OM:93} one may write down coupled 
master equations for propagation forwards and backwards along the $z$ axis
\begin{eqnarray}
 Z_{\pm}(z,t) & = & [1-(\zeta_+ +\zeta_-)\Delta t]Z_{\pm}(z\mp\Delta z,t-\Delta t) \nonumber \\
  && + \zeta_{\mp}\Delta t\bar{Z}_{\pm}(z\mp\Delta z,t+\Delta t)
   + \zeta_{\pm}\Delta tZ_{\mp}(z\pm\Delta z,t-\Delta t)\label{eq:Zpmdef} \\
 \bar{Z}_{\mp}(z\pm\Delta z,t+\Delta t) & = & [1-(\zeta_++\zeta_-)\Delta t]\bar{Z}_{\mp}(z,t)
  + \zeta_{\mp}\Delta tZ_{\mp}(z,t)+\zeta_{\pm}\Delta t\bar{Z}_{\pm}(z,t)\label{eq:Zbpmdef} \\
 Z_{\pm}(z,t) & = & \bar{Z}_{\mp}(z\pm\Delta z,t+\Delta t),\label{eq:ZZBcnst}
\end{eqnarray}
where $Z_{\pm}$ is the forward time propagation probability towards larger $z$, \textit{i.e.}, $Z_+$, or
smaller $z$, \textit{i.e.}, $Z_-$, $\bar{Z}$ is the backward time propagation probability, and
Equation~\ref{eq:ZZBcnst} is the causality constraint. In addition, $\zeta_{\pm}$ is the transition rate
for propagation towards larger ($+$) or smaller ($-$) values of $z$. McKeon and Ord\cite{MO:92} note
that iteration of coupled equations can be difficult. Instead, they inferred a differential equation from
a short time expansion of Equations~\ref{eq:Zpmdef} and~\ref{eq:Zbpmdef} subject to the constraint
Equation~\ref{eq:ZZBcnst}. Performing a Taylor series expansion in $\Delta z = v\Delta t$ and retaining
terms only to order $\Delta t$, one finds the following differential equation for the difference of
$Z_{\pm}$ and $\bar{Z}_{\mp}$
\begin{equation}
 \pm v\left[\frac{\partial Z_{\pm}}{\partial z}-\frac{\partial\bar{Z}_{\mp}}{\partial z}\right]
  +\left[\frac{\partial Z_{\pm}}{\partial t}-\frac{\partial\bar{Z}_{\mp}}{\partial t}\right]
  +\left[\zeta_++\zeta_-\right]\left[Z_{\pm}-\bar{Z}_{\mp}\right]
  = \left[\zeta_{\pm}-\zeta_{\mp}\right]\left[Z_{\mp}-\bar{Z}_{\pm}\right]
  \label{eq:ZZBdif}
\end{equation}
Define $A_{\pm}(z,t) = \exp\left(\left[\zeta_++\zeta_-\right]t\right)(Z_{\pm}(z,t)-\bar{Z}_{\mp}(z,t))$.
One can interpret the integrating factor $\exp\left(\left[\zeta_++\zeta_-\right]t\right)$ as a chemical 
activity, which controls the `concentration' of `up' and `down' transitions. Substituting the definition
of $A_{\pm}$ into Equation~\ref{eq:ZZBdif}, one finds
\begin{equation}
 \pm v\frac{\partial A_{\pm}}{\partial z} + \frac{\partial A_{\pm}}{\partial t} = \left(\zeta_{\pm}-
\zeta_{\mp}\right)A_{\pm}\label{eq:AEOM}
\end{equation}
In order to make further progress, it is useful to Fourier transform Equation~\ref{eq:AEOM} to eliminate
the $z$ and $t$ derivatives and work in the energy momentum representation. Set $v=c$ and define
Fourier amplitudes as follows
\begin{eqnarray}
 A & \equiv & \sum_{p,E}\exp(-i(pz-Et)/\hbar)a_{\pm}(p,E) \label{eq:adef} \\
   & \equiv & \sum_{p,E}\exp(-i(pz+Et)/\hbar)\bar{a}_{\pm}(p,E) \label{eq:abdef}
\end{eqnarray}
Substituting Equations~\ref{eq:adef} and~\ref{eq:abdef} into Equation~\ref{eq:AEOM} and noting that if
the result is to hold for all times over all values of $z$, then the coefficients $a_{\pm}(p,E)$ and
$\bar{a}_{\pm}(p,E)$ must satisfy the following constraints
\begin{eqnarray}
 \mp icpa_{\pm}+iEa_{\pm} & = & \hbar\left(\zeta_{\pm}-\zeta_{\mp}\right)a_{\mp}\label{eq:aEOM}\\
 \mp icp\bar{a}_{\pm}-iE\bar{a}_{\pm} & = & \hbar\left(\zeta_{\pm}-\zeta_{\mp}\right)
  \bar{a}_{\mp}\label{eq:abEOM}
\end{eqnarray}
One may allow $p$ and $E$ to depend on $z$ and $t$ to account for space and time varying potentials, as
will be shown. To be consistent in the order of $\Delta t$ retained, it is sufficient to retain only
the first term in derivatives of $\exp(-i(pz\pm Et))$ with respect to $z$ and $t$, \textit{i.e.}, terms
in $\partial p/\partial z$, $\partial p/\partial t$, $\partial E/\partial z$ and $\partial E/\partial t$
and higher are dropped in the limit $\Delta t\rightarrow 0$.

\section{Interpreting the Equations}

The forward time amplitudes $a_{\pm}(p,E)$ and the time reversed amplitudes $\bar{a}_{\mp}$ encode the
symmetry of the poset used in Knuth and Bahrenyi\cite{KB:09}. Note that
$\left(\zeta_{\pm}-\zeta_{\mp}\right)$ has the same sign as $v$ in Equation~\ref{eq:AEOM}. One may
therefore rewrite $\left(\zeta_{\pm}-\zeta_{\mp}\right)\equiv \pm\omega$ where $\omega$ is a positive
definite constant. One may obtain a dimensionally consistent equation by defining $\hbar\omega = mc^2$,
the Compton energy of the particle. Jacobson and Schulman\cite{JS:84} have an insightful physical
interpretation of mass as being proportional to a rate of making path reversals. That concept seems to
be extended here to a notion of inertia, where the probability of making a particular class of path
reversal $\zeta_+$ or $\zeta_-$ depends on which direction the particle is traveling in.

Multiplying Equations~\ref{eq:aEOM} and~\ref{eq:abEOM} through by $i$ and setting $c=1$, one may write
down matrix equations for $a_{\pm}$ and $\bar{a}_{\pm}$ as follows
\begin{eqnarray}
 E\left(\begin{array}{c}
         a_+ \\
	 a_-
        \end{array}
\right) & = & \left[p\sigma_z+m\sigma_y\right]\left(\begin{array}{c}
         a_+ \\
	 a_-
        \end{array}
\right)\label{eq:asp}\\
 E\left(\begin{array}{c}
         \bar{a}_+ \\
	 \bar{a}_-
        \end{array}
\right) & = & -\left[p\sigma_z+m\sigma_y\right]\left(\begin{array}{c}
         \bar{a}_+ \\
	 \bar{a}_-
        \end{array}
\right)\label{eq:absp}
\end{eqnarray}
It is a straightforward exercise to show that the Equations~\ref{eq:asp} and~\ref{eq:absp} satisfy the
relativistic dispersion relation $E^2 = p^2 + m^2$ in a system of units where $c=1$.

Using the transformation
\begin{displaymath}
 \Phi=\bar{\Phi}^{-1}=\left[\begin{array}{cc}
                             e^{-i\phi/2} & 0 \\
			     0 & e^{i\phi/2}
                            \end{array}
\right]
\end{displaymath}
it is possible to rewrite the equations for $\alpha_{\pm} = a_{\pm}e^{\mp i3\pi/4}$ and 
$\bar{\alpha}_{\pm}=\bar{a}_{\pm}e^{\pm i3\pi/4}$ in a convenient matrix form
\begin{equation}
 \left[\begin{array}{cccc}
         p & -m &  0 &  0 \\
	-m & -p &  0 &  0 \\
	 0 &  0 & -p & -m \\
	 0 &  0 & -m &  p
       \end{array}
\right]\left[\begin{array}{c}
              \alpha_+ \\
	      \alpha_- \\
	      \bar{\alpha}_+\\
	      \bar{\alpha}_-
             \end{array}
\right]=E\left[\begin{array}{c}
              \alpha_+ \\
	      \alpha_- \\
	      \bar{\alpha}_+\\
	      \bar{\alpha}_-
             \end{array}\right]\label{eq:Dintmdt}
\end{equation}
subject to the constraint $E^2 = p^2 + m^2$. Equation~\ref{eq:Dintmdt} is an eigenvalue equation in the
form $H\Psi = E\Psi$. Focusing on $H$, one may apply the following symmetry transformation
\begin{displaymath}
 \Sigma = \Sigma^{-1} = \left[\begin{array}{cccc}
                            1 & 0 & 0 & 0 \\
			    0 & 0 & 1 & 0 \\
			    0 & 1 & 0 & 0 \\
			    0 & 0 & 0 & 1
                           \end{array}
\right]
\end{displaymath}
to put $H$ into the following form
\begin{equation}
 H = \left[\begin{array}{cccc}
             p &  0 & -m &  0 \\
	     0 & -p &  0 & -m \\
	    -m &  0 & -p &  0 \\
	     0 & -m &  0 &  p
           \end{array}
\right]\label{eq:Hpdef}
\end{equation}
Note that Equation~\ref{eq:Hpdef} is in the form
\begin{displaymath}
 H = \left[\begin{array}{cc}
             P & -M \\
	    -M & -P
           \end{array}
\right]
\end{displaymath}
Observe that $M$ is proportional to the unit matrix and is invariant under unitary transformations.
Define
\begin{displaymath}
 P\rightarrow P'' = \left[\begin{array}{cc}
            p_z'' &    0 \\
	      0   & -p_z''
           \end{array}
\right]
\end{displaymath}
and consider a rotation about the $y''$ axis of the form
\begin{displaymath}
 U_{\theta} = U_{\theta}^{-1} = \left[\begin{array}{cc}
                     \cos\theta/2 &  \sin\theta/2 \\
		     \sin\theta/2 & -\cos\theta/2
                    \end{array}
\right]
\end{displaymath}
such that $\tan\theta = p_x'/p_z'$. In this new coordinate system
\begin{displaymath}
 P''\stackrel{\theta}{\rightarrow}P' = \left[\begin{array}{cc}
                                              p_z' &  p_x' \\
					      p_x' & -p_z'
                                             \end{array}
\right]
\end{displaymath}
Now perform a rotation around the $z'$ axis of the form
\begin{displaymath}
 U_{\varphi} = \left(U_{\varphi}^{-1}\right)^*
	     = \left[\begin{array}{cc}
	              e^{-i\varphi/2} & 0 \\
		      0 & e^{i\varphi/2}
	             \end{array}
\right]
\end{displaymath}
such that $\tan\varphi\equiv p_y/p_x$.  In this new coordinate system $p_z'=p_z$. Thus
\begin{displaymath}
 P''\stackrel{\theta}{\rightarrow}P'\stackrel{\varphi}{\rightarrow}P
=\left[\begin{array}{cc}
        p_z      &  p_x-ip_y \\
	p_x+ip_y & -p_z
       \end{array}
\right]
\end{displaymath}
which properly encodes the operator $\mbox{\boldmath $\sigma$}\cdot\mathbf{p}$. The following orthogonal
transformation ($R^T$ indicates the matrix transpose of $R$)
\begin{displaymath}
 R = \left(R^{-1}\right)^T = \frac{1}{\sqrt{2}}\left[\begin{array}{cc}
                                                   1 & -1 \\
						   1 &  1
                                                  \end{array}
\right]
\end{displaymath}
allows one to write
\begin{displaymath}
 RHR^T = \left[\begin{array}{cc}
                M & \mbox{\boldmath $\sigma$}\cdot\mathbf{p} \\
		\mbox{\boldmath $\sigma$}\cdot\mathbf{p} & -M
               \end{array}
\right]
\end{displaymath}
which is equivalent to Dirac's time-independent equation in momentum space
\begin{equation}
 \left[\mbox{\boldmath $\alpha$}\cdot\mathbf{p} + \beta m\right]\psi = E\psi\label{eq:DIRAC}
\end{equation}
where the $\alpha$ and $\beta$ matrices are in the Dirac representation\cite{BD:64}. Note that the
components of $\psi$ are linear combinations of the Fourier amplitudes defined in Equations~\ref{eq:asp} 
and~\ref{eq:absp}. The Fourier amplitudes may be traced back, ultimately, to the master equations for 
the forward and backward transition probability amplitudes defined in Equations~\ref{eq:Zpmdef} 
and~\ref{eq:Zbpmdef} as all of the transformations are invertible. Given that the various $\alpha$
quantities are constantly being rephased and formed into new linear combinations, it is clear that
they must be probability amplitudes and not probabilities \textit{per se}. The picture that emerges
is that a particle follows a stochastic trajectory in time and position, subject to the causality
constraint introduced above. Manipulations of the relevant equations are facilitated by working in the
momentum-energy representation.

\section{Incorporating a Potential}

Given the assumption that $p$ and $E$ may be functions of position and time, one may reinterpret the
Fourier coefficients as canonical momenta. In this way, one would incorporate an 
electromagnetic
four-vector as follows
\begin{displaymath}
 p^{r}\rightarrow p^{r}-\frac{eA^{r}}{c},\quad E\rightarrow E-eA^0/c,
\end{displaymath}
where $r\in\{1,2,3\} $.

Under the assumption that
\begin{displaymath}
 p = \sqrt{(p_x-eA_x/c)^2+(p_y-eA_y/c)^2+(p_z-eA_z/c)^2}
\end{displaymath}
one may use the same series of steps used to derive Equation~\ref{eq:DIRAC} to show that
\begin{displaymath}
 RHR^T = \left[\begin{array}{cc}
                M & \mbox{\boldmath $\sigma$}\cdot(\mathbf{p}-e\mathbf{A}) \\
		\mbox{\boldmath $\sigma$}\cdot(\mathbf{p}-e\mathbf{A}) & -M
               \end{array}
\right]
\end{displaymath}
where $c=1$. Note that, expressed as a four-gradient, the four-momentum transforms as a covariant
(lowered index) four-vector.  With this observation the Dirac equation becomes
\begin{equation}
 \left[\mbox{\boldmath $\alpha$}\cdot(\mathbf{p}-e\mathbf{A}) + \beta m\right]\psi = 
(E-eA_0)\psi\label{eq:DIRACPOT}
\end{equation}
One can recover the standard Dirac equation in the space time representation by substituting
\begin{displaymath}
 \psi=\sum_{\mathbf{p},t}\exp(i(\mathbf{p}\cdot\mathbf{x}-Et))\Psi(\mathbf{p},t)
\end{displaymath}
in Equation~\ref{eq:DIRACPOT}. Note the argument of the exponent is a scalar, so if it is valid in one
frame of reference, then it is valid in all reference frames. Then for $\psi(\mathbf{x},t)$ to be valid
over all space and time, one requires
\begin{equation}
 \left[\mbox{\boldmath $\alpha$}\cdot(-i\mbox{\boldmath $\nabla$}-e\mathbf{A}) + \beta m\right]\psi(x,t)
 = \left(i\frac{\partial}{\partial t}-eA_0\right)\psi(x,t)\label{eq:DIRACPOSTIM}
\end{equation}
Equation~\ref{eq:DIRACPOSTIM} completes the derivation of the time-dependent Dirac equation in the
presence of a potential from the master equation approach of McKeon and Ord\cite{MO:92}.  In Feynman
slash notation, one has
\begin{displaymath}
 (\slashed{p} -\slashed{A} + m)\psi = 0.
\end{displaymath}

The derivation given here is fairly simple and straightforward compared to other approaches to the Dirac
equation in `$3+1$' dimensions based on the Feynman checkerboard
problem\cite[and references therein]{S:05}. This may be due to the observation
that the only essential property of spinors used here is that the quadratic expression
$p_x^2 + p_y^2 + p_z^2 = p^2$ can be `bilinearized' into the form
\begin{displaymath}
 \left[\begin{array}{cc}
        p_z & p_x-ip_y \\
	p_x+ip_y & -p_z
       \end{array}
 \right]
 \left[\begin{array}{cc}
        p_z & p_x-ip_y \\
	p_x+ip_y & -p_z
       \end{array}
 \right] = \left(p_x^2 + p_y^2 + p_z^2\right)
 \left[\begin{array}{cc}
        1 & 0 \\
	0 & 1
       \end{array}
 \right],
\end{displaymath}
a result which has much more to do with analytic geometry than any notions of `quantum strangeness'.
In order to complete the program sketched in the abstract, it is necessary to show how the master
equation may be inferred from the poset approach of Knuth and Bahrenyi\cite{KB:09}. As noted, this is
work in progress. 
\vfill
\eject
\section*{Acknowledgements}

KAE thanks the University at Albany for partial support of this work through its Faculty Research Award
Program. KAE also thanks ACERT (NIH NCRR P41 RR016292) for the use of its computational resources.

\vfill
\eject

\bibliography{Earle_docs,Intrinsic,%
Analytical,Dirac,Statistical,DiracEquation}

\begin{thebibliography}{10}

\bibitem{FH:65}
Richard~P. Feynman and A.~R. Hibbs.
\newblock {\em {Quantum Mechanics and Integrals}}.
\newblock McGraw-Hill, 1965.

\bibitem{S:05}
L.~S. Schulman.
\newblock {\em {Techniques and Applications of Path Integration}}.
\newblock Dover, 2005.

\bibitem{BD:64}
J.~D. Bjorken and S.~D. Drell.
\newblock {\em {Relativistic Quantum Mechanics}}.
\newblock {McGraw-Hill}, 1964.

\bibitem{JS:84}
Theodore Jacobson and Lawrence~S. Schulman.
\newblock {Quantum stochastics: the passage from a relativistic to a
  non-relativistic path integral}.
\newblock {\em J. Phys. A: Math. Gen.}, 17:375--383, 1984.

\bibitem{E:10}
K.~A. Earle.
\newblock {Notes on the Feynman Checkerboard Problem}.
\newblock arXiv:1012.1564v1 [math-ph], 2010.

\bibitem{G:81}
H.~A. Gersch.
\newblock {Feynman's Relativistic Chessboard as an Ising Model}.
\newblock {\em International Journal of Theoretical Physics}, 20:491--501,
  1981.

\bibitem{O:92a}
G.~N. Ord.
\newblock {A Reformulation of the Feynman Chessboard Model}.
\newblock {\em Journal of Statistical Physics}, 66:647--659, 1992.

\bibitem{O:92b}
G.~N. Ord.
\newblock {Classical Analogue of Quantum Phase}.
\newblock {\em International Journal of Theoretical Physics}, 31:1177--1195,
  1992.

\bibitem{MO:92}
D.~G.~C. McKeon and G.~N. Ord.
\newblock {Time Reversal in Stochastic Processes and the Dirac Equation}.
\newblock {\em Physical Review Letters}, 69:3--4, 1992.

\bibitem{OM:93}
G.~N. Ord and D.~G.~C. McKeon.
\newblock {On the Dirac Equation in 3+1 Dimensions}.
\newblock {\em Annals of Physics}, 222:244--253, 1993.

\bibitem{K:90}
A.~L. Kholodenko.
\newblock {Fermi-Bose Transmutation: From Semiflexible Polymers to
  Superstrings}.
\newblock {\em Annals of Physics}, 202:186--225, 1990.

\bibitem{K:91}
A.~L. Kholodenko.
\newblock {Topological Theory of Reptation}.
\newblock {\em Physics Letters A}, 159:437--441, 1991.

\bibitem{S:68}
L.~S. Schulman.
\newblock {A Path Integral for Spin}.
\newblock {\em Physical Review}, 176:1558--1569, 1968.

\bibitem{KB:09}
K.~H. Knuth and N.~Bahrenyi.
\newblock {A Derivation of Special Relativity from Causal Sets}.
\newblock arXiv:1005.4172v2 [math-ph], 2010.

\bibitem{DP:02}
B.~A. Davey and H.~A. Priestly.
\newblock {\em {Introduction to Lattices and Order}}.
\newblock Cambridge University Press, 2002.

\end{thebibliography}
\bibliographystyle{unsrt}

\end{document}